\documentclass[pre, twocolumn]{revtex4-1} 
\usepackage{amsmath}  
\usepackage{amsfonts} 
\usepackage{graphicx} 
\usepackage{subfigure}

\begin{document}
\bibliographystyle{unsrt}
\title{Effective ergodicity in single-spin-flip dynamics}
\author{Mehmet S\"uzen}
\affiliation{Applied Mathematical Physiology Lab \\ Bonn University, Sigmund-Freud-Str.25, 53127 Bonn, Germany} 
\email{mehmet.suzen@physics.org}
\date{\today}
\begin{abstract}
A quantitative measure of convergence to effective ergodicity, Thirumalai-Mountain (TM) metric, is applied to Metropolis 
and Glauber single spin flip dynamics. In computing this measure, finite lattice ensemble averages are obtained using the exact 
solution for one dimensional Ising model, where as, the time averages are computed with Monte 
Carlo simulations. The time evolution of effective ergodic convergence of Ising magnetization 
is monitored. By this approach, diffusion regimes of the effective ergodic convergence 
of magnetization are identified for different lattice sizes, non-zero temperature and non-zero external 
field values. Results show that caution should be taken using TM metric at system parameters
that give rise to strong correlations.
\end{abstract}
\maketitle


\section{Introduction}

Cooperative phenomena are present in many different fields \cite{wannier45a}. 
A unifying approach in studying cooperation among individual units emerge 
as a mathematical model that most resembles the nature of the problem. 
The first and the most successful of these models which were exactly solvable 
was the one dimensional Ising model, a closed chain of $n$ cooperating units, 
mimicking spins in ferromagnetic materials \cite{ising25a, brush67a, baxter82a}. 
Time dependent statistics of the Ising model has been studied in depth before \cite{glauber63a, binder2010monte}, 
where single flip dynamics on $n$ spins is introduced by changing a single spin's value 
with an associated transition probability. Natural consequence of generating such dynamics 
in a given statistical ensemble is the question of how and 
when the system behaves ergodically, i.e., ensemble averages being equivalent 
to the time averages. This question is not only interesting due to fulfilling 
Boltzmann's equilibrium statistical mechanics \cite{tolman, farquhar, dorfman99a}, 
but for its crucial importance in practical applications, such as, in simple liquids 
\cite{mountain89me, de2005diagnosing}, assessing quality of the Monte Carlo simulations 
\cite{neirotti2000approach}, earthquake fault networks \cite{tiampo03a, tiampo2007a} and 
in econophysics \cite{peters11a}. Most of these studies address 
the problem of identifying ergodic or non-ergodic regimes.
In this study, we investigate the time evolution of the rate of effective 
ergodic convergence under different system parameters to identify its so called 
diffusion regimes.  

The Ising model and its analytic solution for the finite size total magnetization 
corresponding to the ensemble average are introduced in Sec. \ref{sec:is}. In 
Sec. \ref{sec:mc} we will provide details of our strategy of computing time 
averages using Metropolis and Glauber single spin dynamics defined on the Ising model.
In Sec. \ref{sec:ergo} we briefly review the basic definitions of ergodicity from applied 
statistical mechanics point of view.  The mathematical literature based on measure 
theory is largely ignored. However, a quantitative measure for the identification 
of effective ergodic dynamic is needed. In Sec. \ref{sec:tm} the fluctuation metric 
\cite{mountain89me, thirumalai1989ergodic} is adapted for Ising model's total 
magnetization. By this approach, the rate of effective ergodic convergence of 
magnetization is monitored in single spin flip dynamics. We report the diffusion 
behavior of the ergodic convergence and identify different regimes depending on 
different lattice sizes, temperature and external field values in Sec. \ref{sec:diff}.

\section{The Ising model \label{sec:is}}

Consider a one dimensional lattice that contains $N$ sites. Each site's value can be 
labelled as $\{s_{i}\}_{i=1}^{N}$. In the two state version of the lattice, which is 
the Ising model  \cite{ising25a, brush67a, baxter82a}, sites can take up two values, 
such as $\{1,-1\}$. These values correspond to spin up and spin down states, 
for example as a model of magnetic material or the state of a neuron \cite{hopfield1982}.

The total energy, Hamiltonian of the system can be written as follows

\begin{eqnarray}
 \label{eq:Ham1}
  \mathcal{H}(\{s_{i}\}_{i=1}^{N}, J, H) & = & J \Big( (\sum_{i=1}^{N-1} s_{i} s_{i+1}) + (s_{1} s_{N}) \Big)  \nonumber \\
                                         &   & + H \sum_{1}^{N} s_{i}.
\end{eqnarray}

This expression contains two interactions, one due to nearest-neighbors (NN) and one due to 
an external field. Note that, additional term in NN interactions $s_{1} s_{N}$ term appears 
due to periodic or cyclic boundary conditions to provide translational invariance.
Coefficients $J$ and $H$ corresponds to scaling of these interactions 
respectively. A reduced form is used in Eq.(\ref{eq:Ham1}) using the unit thermal 
background, using the Boltzmann factor $\beta = \frac{1}{k_{B}T}$,

\begin{eqnarray}
   K = \beta J, \qquad h=\beta H.
\end{eqnarray}

The partition function for this system can be written by using the transfer matrix technique \cite{baxter82a}
\begin{eqnarray}
  \label{eq:Zn}
  Z_{N} = Tr \big( V^N \big).
\end{eqnarray}
$V$ is the transfer matrix defined as follows
\begin{eqnarray}
  \label{eq:matrix}
  V = 
  \begin{pmatrix}
   e^{K+h} & e^{-K}  \\
   e^{-K} & e^{K-h}  
  \end{pmatrix}.
\end{eqnarray}
The resulting free energy for the finite system appears in terms of eigenvalues of the transfer matrix, $\lambda_{1}$ 
and $\lambda_{2}$ \cite{baxter82a}
\begin{eqnarray}
  \label{eq:eigenZ}
   Z_{N}        & = & \lambda_{1}^N + \lambda_{2}^N,  \\
  \lambda_{1,2} & = & e^{K} \big[ \cosh{h} \pm \sqrt{\sin^2 h + e^{-4K}} \big].
\end{eqnarray}
The canonical free energy for the finite system is defined as follows \cite{baxter82a}
\begin{eqnarray}
  \label{eq:eigenZ}
    f(N, T, h)         & = & - k_{B} T \frac{1}{N} ln Z_{N}, \\
  \frac{1}{N} ln Z_{N} & = & ln \lambda_{1} +  \frac{1}{N} ln \big[ 1 + (\lambda_{2}/\lambda_{1})^{N} \big].
\end{eqnarray}

We are interested in {\it finite size} total magnetization, to compute the ensemble average of it, $M_{E}(N, \beta, H)$,
analytically. Differentiating canonical free energy with respect to $H$ will yield a long expression for $M_{E}$,
\begin{eqnarray}
  \label{eq:Me}
  M_{E}( N, \beta, H) & = & \big(N M_{1} M_{2}^{N-1} + N M_{3} M_{4}^{N-1} \big) / M_{5}, \\ 
                 M_{1}& = & {{\beta \,\cosh \left(H\, \beta \right)\,\sinh \left(H\,\beta\right)
 }\over{\sqrt{e^ {- 4\, \beta \,J }+\sinh ^2\left(H\,\beta \right)}}} + \beta \,\sinh \left(H\,\beta \right),   \nonumber  \\
                 M_{2}& = & \sqrt{e^ {- 4\,\beta\,J }+\sinh ^2\left( H\,\beta \right)}+\cosh \left(h\,\beta \right), \nonumber \\
                 M_{3} & = & -M_{1} + 2 \beta \,\sinh \left(H\,\beta \right),  \nonumber \\
                 M_{4} & = & -M_{2} + 2 \cosh \left(H\,\beta \right),   \nonumber \\
                 M_{5} & = & \big(\sqrt{e^ {- 4\, \beta \,J }+\sinh ^2\left(H\,\beta \right)} + \cosh \left(H\,\beta \right) \big)^{N} \nonumber \\
                       &  &  + \big( \cosh \left(H\,\beta \right) - \sqrt{e^ {- 4\, \beta \,J }+\sinh ^2\left(H\,\beta \right)} \big)^{N}. \nonumber
\end{eqnarray}

Note that in Eq.(\ref{eq:Me}), the Boltzmann factor is explicitly written. Further explorations
of the analytical solutions are beyond the scope of this study.

\section{Metropolis and Glauber single spin flip dynamics \label{sec:mc}}

One of the ways to generate dynamics for a lattice system similar to Ising model in a computer simulation is changing
the value of a randomly chosen site to its opposite value. This procedure is called {\it single spin flip dynamics} in the 
context of Monte Carlo simulations \cite{binder2010monte}.  However, the quality of this kind of dynamics depends highly 
on the quality of the random number generator (RNG) \cite{compagner91, janke2002pseudo} we employ in selecting the 
site to be flipped. However, we gather that Marsenne-Twister as an RNG \cite{matsumoto98a} is sufficiently good 
for this purpose.

In generating such a dynamics, there is an associated transition probability in the single spin flip. This probability
would determine if the flip introduced by the Monte Carlo procedure is an acceptable physical move. Two forms 
of transition probability can be used that correspond to Boltzmann density.  The following expressions generate
Glauber and Metropolis dynamics respectively,
\begin{eqnarray}
 \label{eq:transP}
p_{Glauber}(\{s_{i}\}_{i=1}^{N})   & = & \exp(-\beta  \Delta \mathcal{H})/ \big( 1 + \exp(-\beta  \Delta \mathcal{H}) \big),  \nonumber \\
                                   & = & 1/\big( 1 + \exp(\beta  \Delta \mathcal{H}) \big), \\
p_{Metropolis}(\{s_{i}\}_{i=1}^{N})& = & min\big(1, \exp(-\beta  \Delta \mathcal{H}) \big).
\end{eqnarray}
where $\Delta \mathcal{H}$ is the total energy difference between single spin flipped and non-flipped configurations. 
The resulting transition probability is compared against a randomly generated number $r$, where $r \in [0, 1]$. The move
is accepted if the transition probability is smaller then $r$. This procedure, generally known as Metropolis-Hastings
Monte Carlo,  samples the canonical ensemble \cite{binder2010monte}. 

\section{Ergodicity \label{sec:ergo}}

Boltzmann made the hypothesis that the solution of any dynamical system,
its trajectories, will evolve in time over phase-space regions where macroscopic 
properties are close to the thermodynamic equilibrium \cite{dorfman99a}. 
Consequently, ensemble averages and time averages will yield
the same measure in thermodynamic equilibrium. A form of this
hypothesis states that average values of an observable $g$ over its
ensemble of accessible state points, namely ensemble averaged value 
can be recovered by time averaged values of the observable's 
time evolution, $g(t)$ from $t_{0}$ to $t_{N}$, 
\begin{eqnarray}
 \label{eq:erEn}
   \langle g \rangle = lim_{t_{N} \to \infty} \int_{t_{0}}^{t_{N}} g(t) dt,
\end{eqnarray}
where $\langle \rangle$ indicates ensemble averaged value. Note that, 
the definition of {\it ergodicity} is not uniform in the 
literature \cite{mastatistical, mountain89me, kingman61}. Some works require 
that system should visit all accessible states in the phase-space to reach ergodic
behavior. This is seldomly true. And considering the fact that coarse graining 
of phase-space occurs, most of the accessible state values are clustered.
Frequently, {\it effective ergodicity} can be reached if the system uniformly samples 
the coarse-grained regions relatively quickly \cite{mountain89me}.  

Conditions of {\it ergodicity} in the transition states, a stochastic matrix of transition probabilities,
generated by {\it spin flip dynamics} is studied in the context of Markov chains \cite{kingman61, pakes69}. This 
type of {\it ergodicity} implies that any state can be reached from any other. The Monte Carlo procedure 
explained above may be ergodic by construction in this sense for long enough times. 

\section{Concept of Ergodic Convergence \label{sec:tm}}

A quantitative measure of effective ergodic convergence relies on the 
fact that identical components of the system, particles or a discrete sites, 
carry identical average characteristics at thermal equilibrium  \cite{mountain89me}.
Hence, effective ergodic convergence, $\Omega_{G}(t)$, can be quantified 
over time for an observable, a property, $g$. Essentially it can be 
computed as a difference the between ensemble averaged value of $g$ and the 
sum of the instantaneous values of $g$ for each of the components. This is termed 
Thirumalai-Mountain (TM) $G$-fluctuating metric \cite{mountain89me, thirumalai1989ergodic}, expressed as follows
at a given time $t_{k}$
\begin{eqnarray}
 \label{eq:Og} 
 \Omega_{G}(t_{k}) = \frac{1}{N} \sum_{j=1}^{N} \big[ g_{j}(t_{k}) - \langle g(t_{k}) \rangle \big]^{2},
\end{eqnarray}

where $g_{j}(t_{k})$ is the time-averaged per component and $\langle g(t_{k}) \rangle$ is the instantaneous ensemble 
average defined as 
\begin{eqnarray}
 \label{eq:gEns}
  g_{j}(t_{k})                 & = & \frac{1}{k} \sum_{i=0}^{k} g_{j}(t_{i}), \\
  \langle g(t_{k}) \rangle     & = & \frac{1}{N} \sum_{j=1}^{N} g_{j}(t_{k}).
\end{eqnarray}
Hence the rate of ergodic convergence is measured with
\begin{eqnarray}
  \label{eq:Dg1}
   \Omega_{G}^{'} = \frac{\Omega_{G}(t)}{\Omega_{G}(0)} \to \frac{1}{tD_{G}}
\end{eqnarray}
where $D_{G}$ is the property's diffusion coefficient and $\Omega_{G}$ effective ergodic convergence. If $1/\Omega_{G}^{'}$ is linear in time, any point in phase-space is said to be equally likely. This approach is used 
in simple liquids \cite{mountain89me, de2005diagnosing}, and earthquake fault networks \cite{tiampo03a, tiampo2007a}. 

We would like to investigate the behavior of $1/\Omega_{G}^{'}$ for the Ising model. The adaption
of the  $\Omega_{G}$ for total magnetization at time $t_{k}$ as a function of temperature 
and external field values reads
\begin{eqnarray}
  \label{eq:magG}
   \Omega_{M} (t_{k}, N, \beta, h)  & = & \big[ M_{T}(t_{k}) - M_{E}\big]^{2}, \\
                             M_{T}  & = & \frac{1}{k} \sum_{i=0}^{k} M(t_{i}), \nonumber \\
\end{eqnarray}
where $M_{T}(N, \beta, h)$ and $M_{E}( N, \beta, h)$ correspond to time and ensemble averaged total magnetization. Note
that the value of $M_{E}( N, \beta, h)$ is fixed and is computed using the analytical solutions given in Sec. \ref{sec:is}, where
as $M_{T}(N, \beta, h)$ is computed in the course of Metropolis or Glauber dynamics. Here we slightly differ in
comparison to the TM approach and use constant ensemble average, because in our case the value of the ensemble average 
is available in exact form as given in Eq.(\ref{eq:Me}). 

\section{Diffusion Regimes \label{sec:diff}}

We have identified the time evolution of the effective ergodic convergence measure, 
$\Omega_{M} (t_{k}, N, \beta, h)$, for the total magnetization of a one dimensional 
Ising model. Depending on which transition probability is used for the acceptance criterion, 
we generate Metropolis and Glauber single spin flip dynamics for the following model parameters: 
number of spin sites $N=\{32, 64, 128, 256, 512\}$, Boltzmann factor $\beta=\{ 0.5, 1.0\}$ and 
non-zero external field values $H=\{0.50, 1.0\}$ with setting short-range interaction
strength to $J=1.0$ for all cases \cite{isingLenzMC}. We generate a dynamics up to half a million Monte Carlo steps
for all combination of parameters, hence the maximum $k$. At the rejected moves,
rejected single spin flip configurations, the value of $\Omega_{M} (t_{k}, N, \beta, h)$ is set to 
the previous accepted value. We did not use external field values and temperatures close to zero, 
because total magnetization's exact solution fails for zero temperature. In the case of zero external 
field, total magnetization is zero and Monte Carlo relaxation time is long.

We have generated a set of time evolutions of the effective ergodic convergence measure combined 
in three different schemes: varying external fields, increasing number of spin sites and
different temperature values. For better statistics, $512$ spin sites are used for the 
variation of external fields and temperatures. By employing such a combination scheme, 
we could judge the relations among the variation of different
parameters in the behavior of the ergodic convergence measure over time. The Monte Carlo
steps play a role of {\it pseudo-dynamical} time.

\begin{figure}[ptb]
  \centering
  \subfigure[\label{fig:glauberVaryNa}]{ \includegraphics[width=0.48\columnwidth]{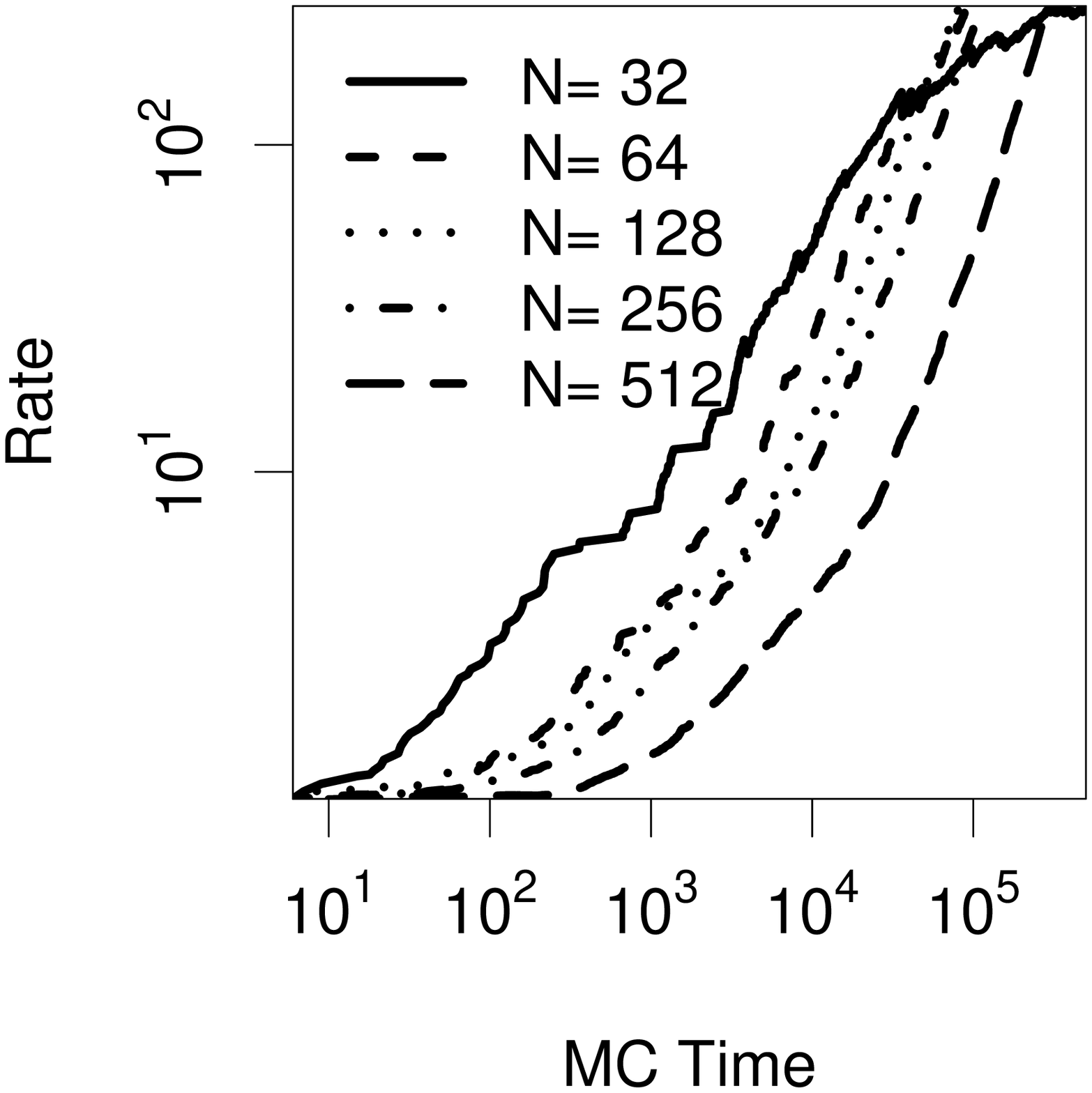}} 
  \subfigure[\label{fig:glauberVaryNb}]{ \includegraphics[width=0.48\columnwidth]{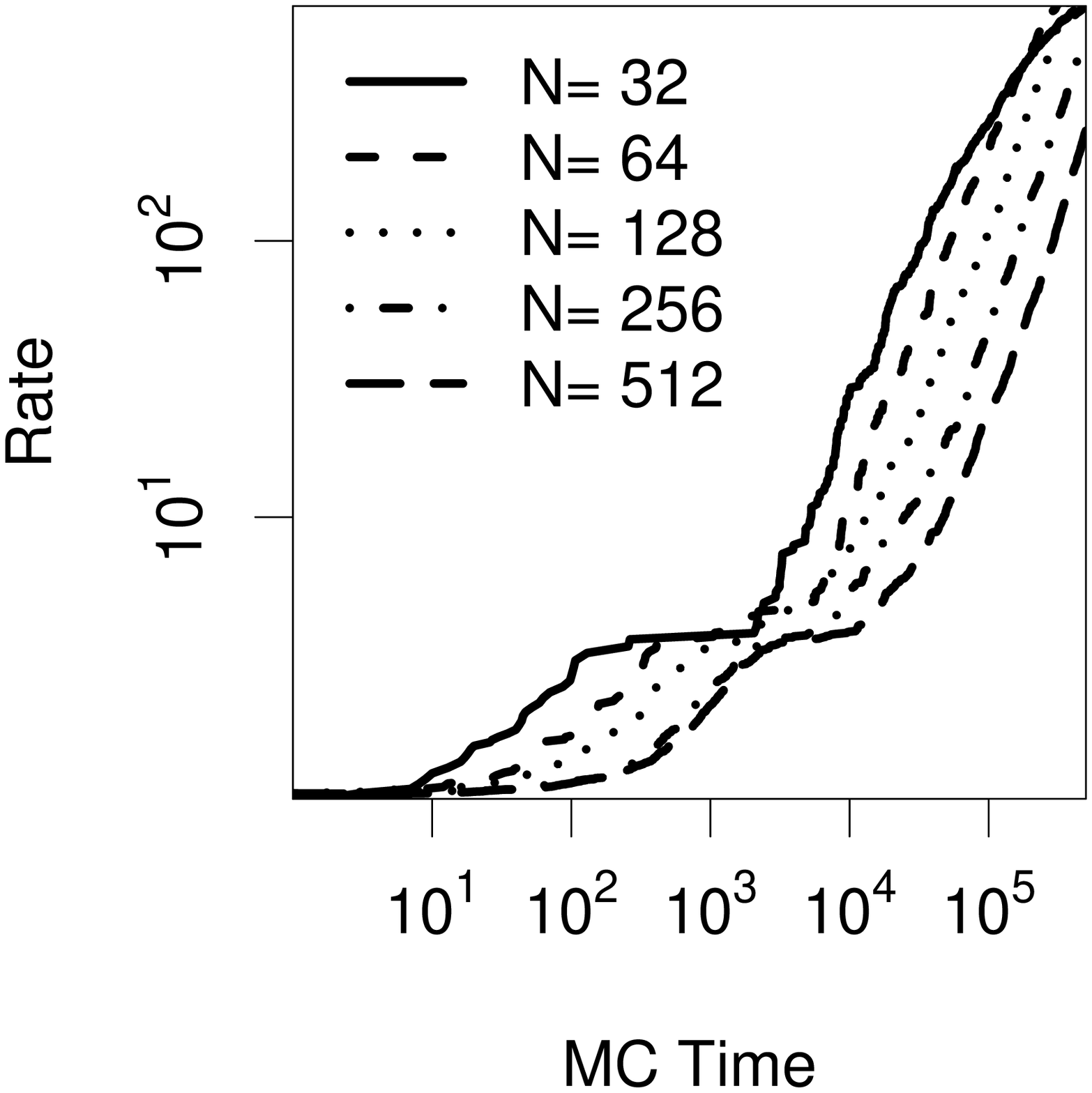}}
  \caption{Inverse effective ergodic convergence in Glauber dynamics with different system sizes with fixed
           temperature $\beta=1.0$ and external fields $H=\{0.5, 1.0\}$ at (a) and (b) respectively.}
\end{figure}

\begin{figure}[ptb]
  \centering
  \subfigure[\label{fig:metropolisVaryNa}]{ \includegraphics[width=0.48\columnwidth]{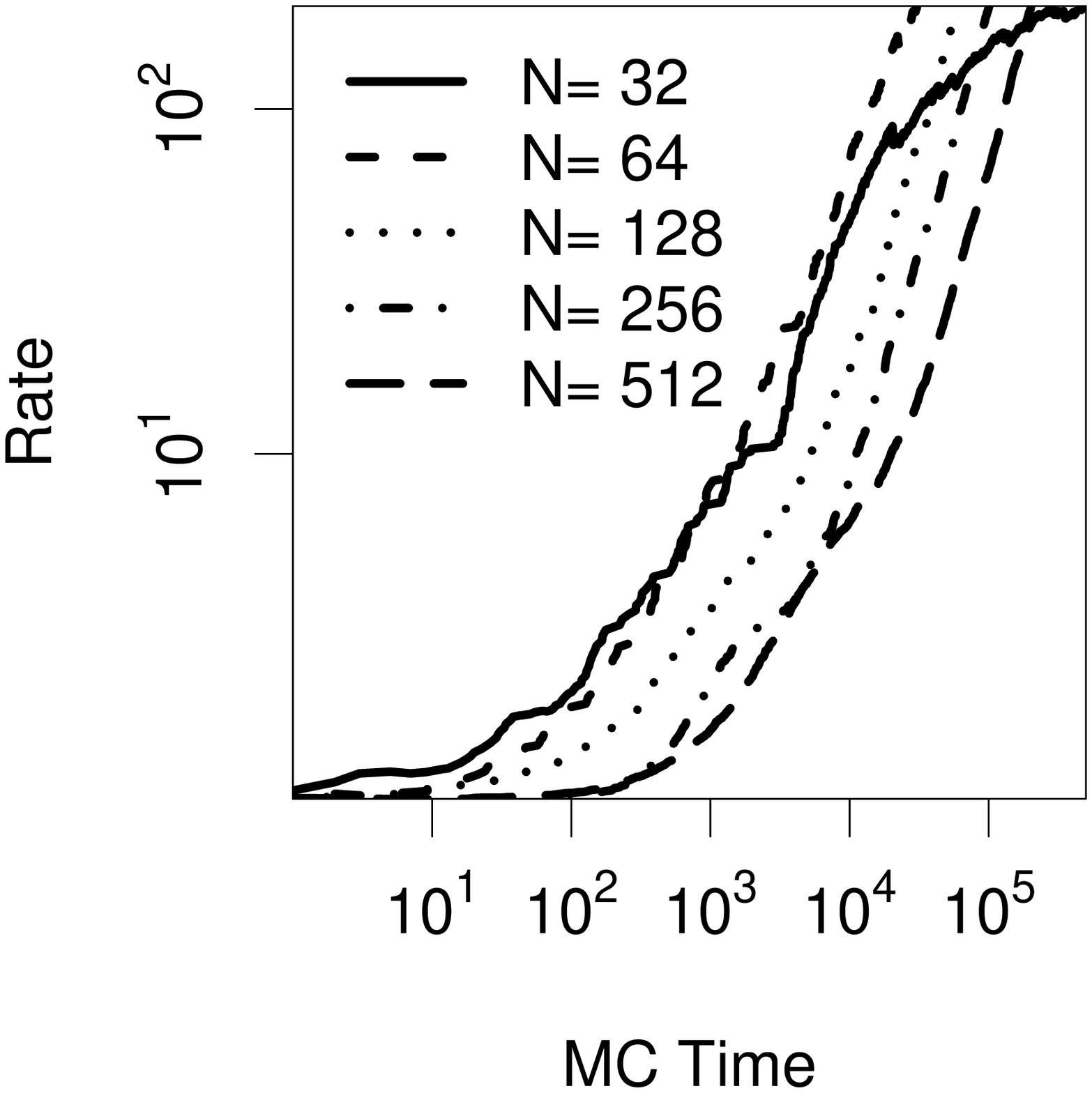}}
  \subfigure[\label{fig:metropolisVaryNb}]{\includegraphics[width=0.48\columnwidth]{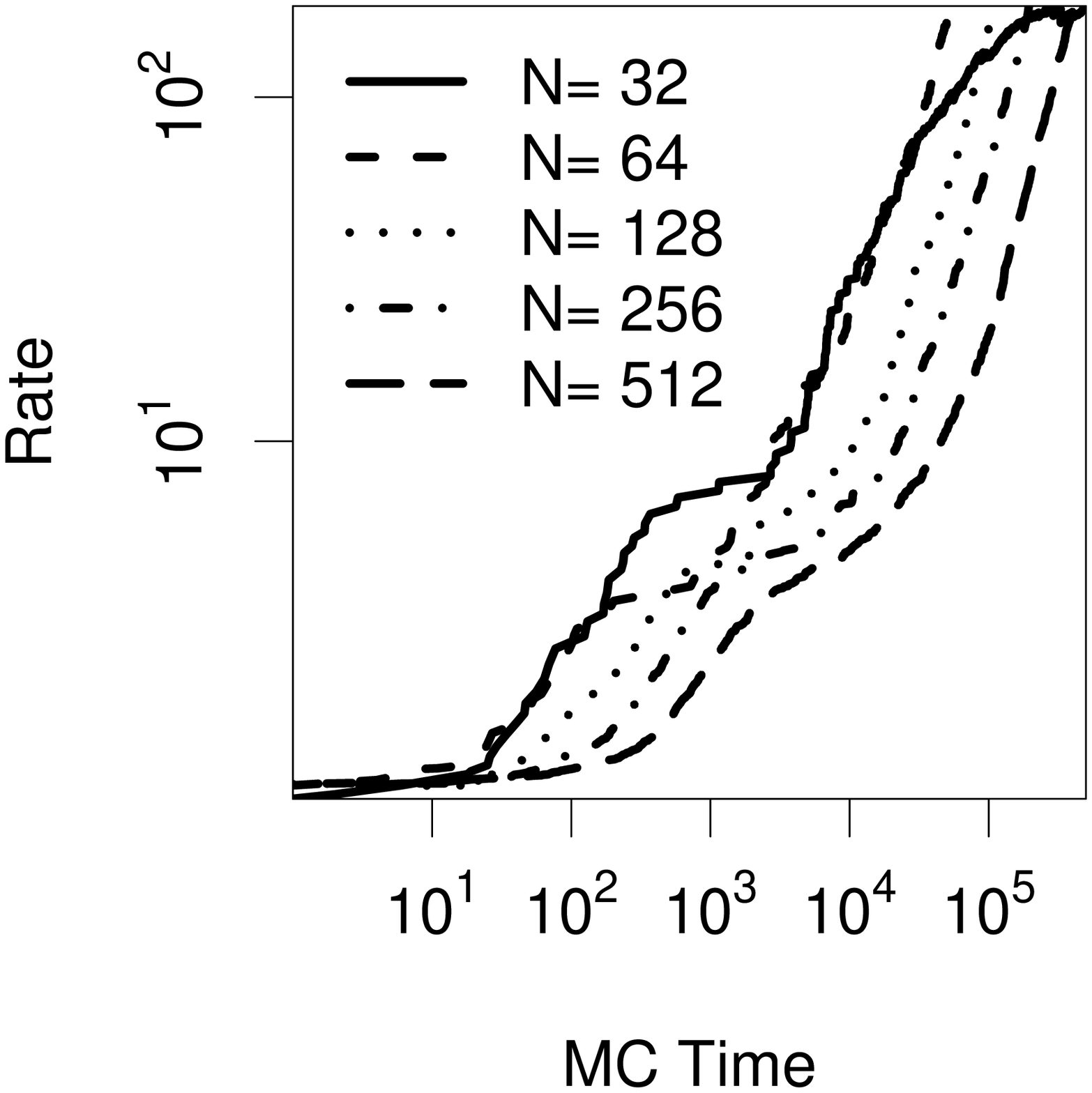}}
  \caption{Inverse effective ergodic convergence in Metropolis dynamics with different system sizes with fixed
           temperature $\beta=1.0$ and external fields $H=\{0.5, 1.0\}$ at (a) and (b) respectively.}
\end{figure}

To be able to judge the diffusion behavior of the  time evolutions of the effective ergodic 
convergence measure, we used the following expression with $D_{M}$, the diffusion coefficient,
\begin{eqnarray}
  \label{eq:invM}
  1/\Omega_{M}^{'} = \frac{\Omega_{M}(t_{0}, N, \beta, h)}{\Omega_{M}(t_{k}, N, \beta, h)} & \to & tD_{M}.
\end{eqnarray}
We call this value so called {\it inverse effective ergodic convergence rate}, simply {\it the rate}.
The rate in our plots shows an increasing value over time. A higher value implies that the system is closer
to ergodic regime.

Figures \ref{fig:glauberVaryNa}, \ref{fig:glauberVaryNb} and Figures \ref{fig:metropolisVaryNa}, \ref{fig:metropolisVaryNb} 
show the effect of the lattice size, 
different number of spin sites, two different external field values at fixed unit thermal background, 
for Glauber and Metropolis dynamics respectively. It is seen in all cases that smaller size leads to 
{\it faster} ergodic convergence. This behavior is more pronounced with the Glauber dynamics. It is well known that 
Glauber dynamics provides faster convergence to equilibrium \cite{binder2010monte}. When the external field is 
higher, at $1.0$, we observe two different diffusion regimes. Those regimes can be clearly judged from 
inflection points given on the rate curves. Those inflection points, plateau regions, are significant in the 
Glauber dynamics. Again, the plateau regions are shifted for smaller size configurations to the left of the 
figure, due to faster convergence we mentioned. 

\begin{figure}[ptb] 
  \centering
  \subfigure[\label{fig:glauberVaryHa}]{\includegraphics[width=0.48\columnwidth]{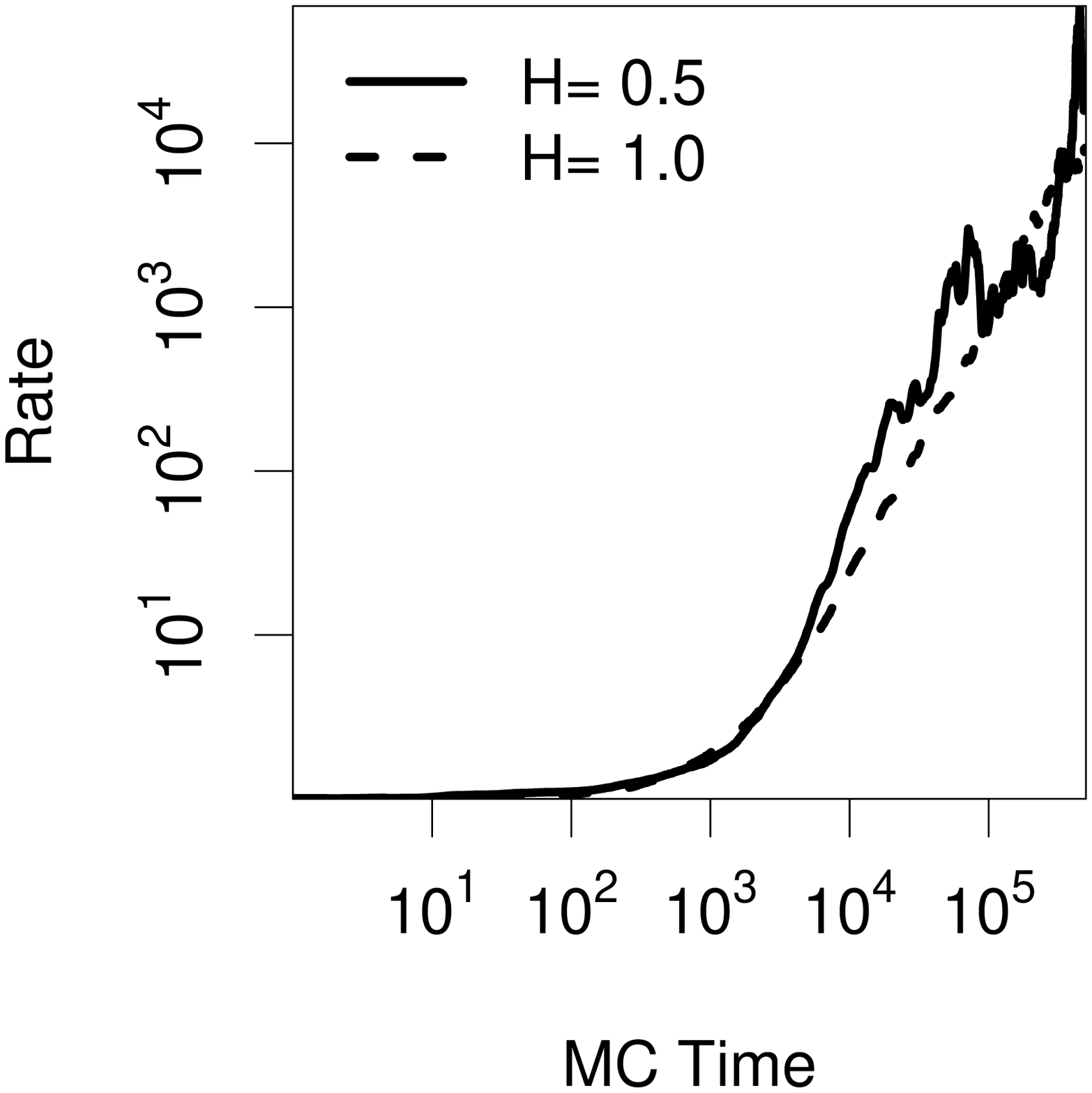}}
  \subfigure[\label{fig:glauberVaryHb}]{\includegraphics[width=0.48\columnwidth]{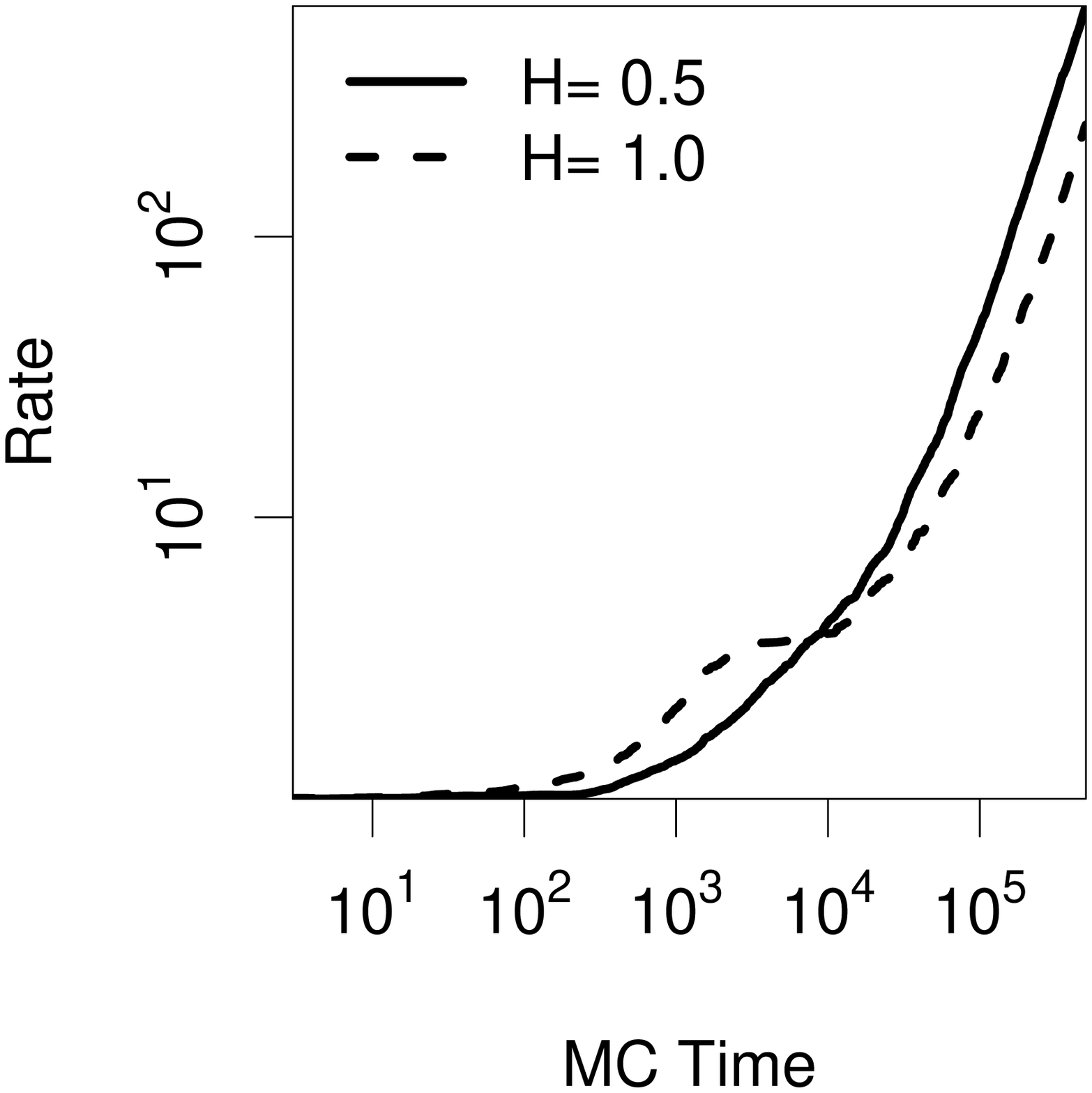}}
  \caption{Effective ergodic convergence in Glauber dynamics with different external field values with fixed 
           size $N=512$ and temperature $\beta=\{0.5, 1.0\}$ at (a) and (b) respectively.}
\end{figure}

For varying external field values, there is only a single diffusion regime for low external field values. 
However upon increasing of the field values we again observe inflection point in the rate curves. This 
signifies two different diffusion regimes for the rate. This is demonstrated in Figure \ref{fig:glauberVaryHa} and 
\ref{fig:glauberVaryHb}. 

\begin{figure}[ptb]
  \centering
  \subfigure[\label{fig:glauberVaryTa}]{\includegraphics[width=0.48\columnwidth]{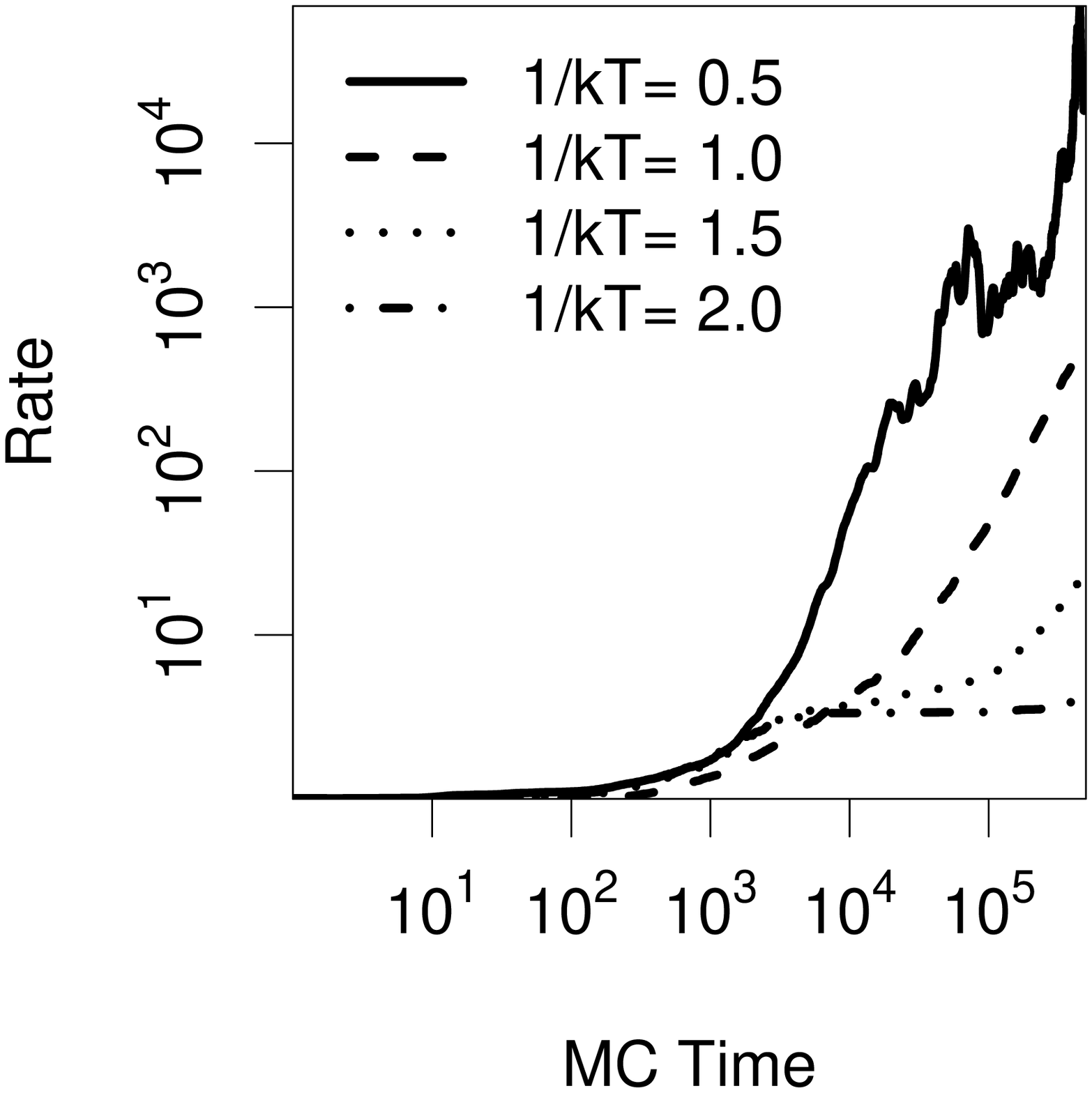}}
  \subfigure[\label{fig:glauberVaryTb}]{\includegraphics[width=0.48\columnwidth]{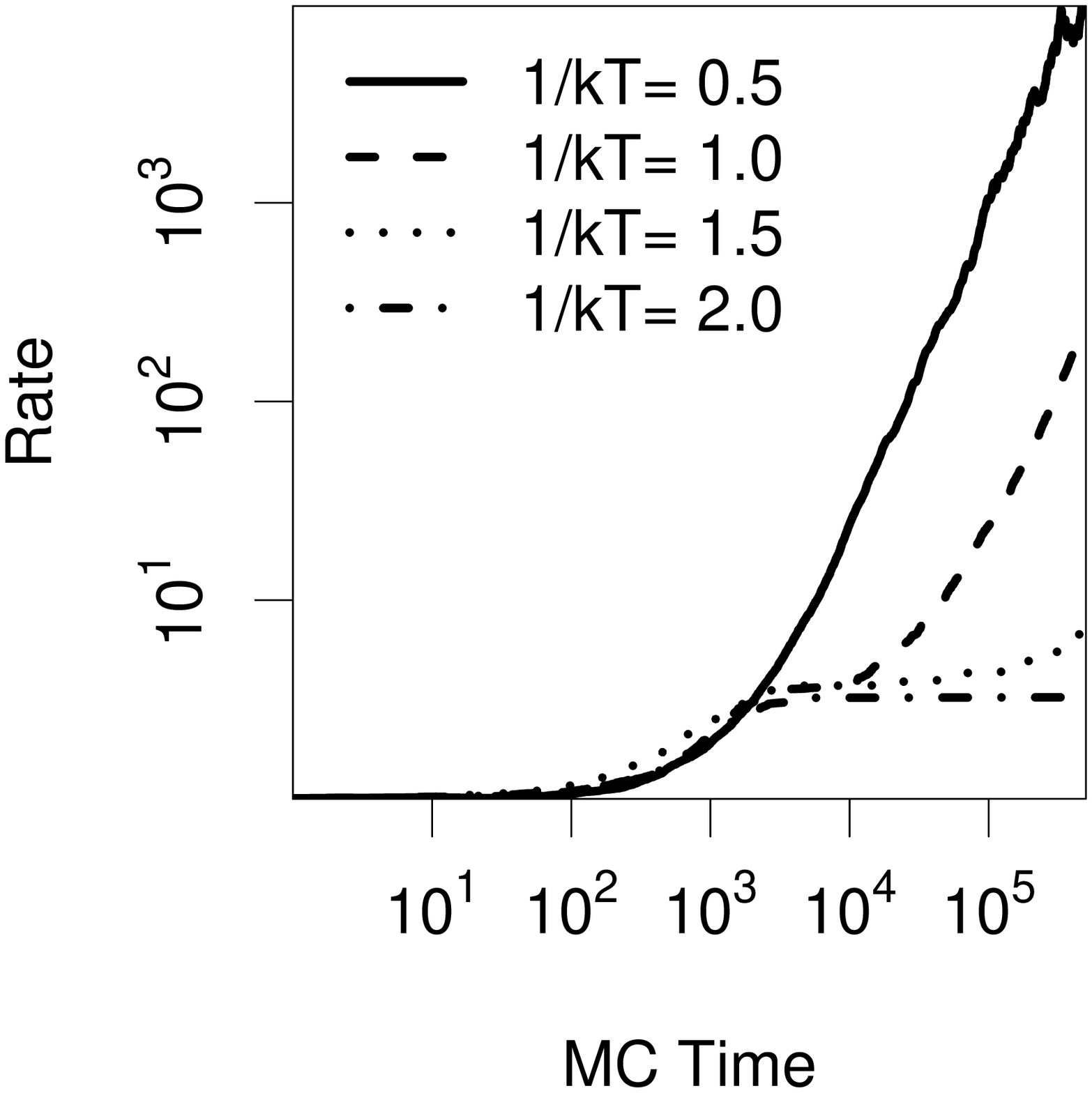}}
  \caption{Effective ergodic convergence in Glauber dynamics with different temperatures with fixed
           size $N=512$ and  external fields $H=\{0.5, 1.0\}$ at (a) and (b) respectively.}
\end{figure}

Temperature dependence of the rate curve is shown in Figure \ref{fig:glauberVaryTa} and \ref{fig:glauberVaryTb}. We see that combination of
higher temperature and external field values induce a change in the diffusion behavior. We observe that
plateau regions become larger upon increasing temperature.

\section{Summary \label{sec:summ}}
The behavior of {\it the rate} of convergence to ergodicity is characterized for the Ising model
using the modified Thirumalai-Mountain (TM) metric for the total magnetization. We aimed at determining 
the rate's behavior over time. We conclude that combination of stronger temperature or external field values generates 
a regime change in the ergodic convergence. Hence, caution should be taken using TM metric at system parameters 
that give rise to strong correlations.

\section*{Acknowledgements}
The author would like to thank Ole Peters for pointing out the original Thirumalai-Mountain metric,
Cornelius Weber, Richard M. Neumann, Ely Klepfish and Iqbal Hussain for critical reading of the 
manuscript and fruitful correspondence, anonymous referees and the editor for helpful comments 
in the review process which have resulted in an improved manuscript and presentation.
\bibliography{lib}

\begin{thebibliography}{10}

\bibitem{wannier45a}
G.~H. Wannier.
\newblock The statistical problem in cooperative phenomena.
\newblock {\em Rev. Mod. Phys.}, 17:50--60, Jan 1945.

\bibitem{ising25a}
Ernst Ising.
\newblock Beitrag zur theorie des ferromagnetismus.
\newblock {\em Zeitschrift f{\"u}r Physik A Hadrons and Nuclei},
  31(1):253--258, 1925.

\bibitem{brush67a}
Stephen~G Brush.
\newblock History of the lenz-ising model.
\newblock {\em Reviews of Modern Physics}, 39(4):883, 1967.

\bibitem{baxter82a}
RJ~Baxter.
\newblock {\em Exactly solvable models in statistical mechanics}.
\newblock Academic Press London, 1982.

\bibitem{glauber63a}
Roy~J Glauber.
\newblock Time-dependent statistics of the ising model.
\newblock {\em Journal of Mathematical Physics}, 4(2):294--307, 1963.

\bibitem{binder2010monte}
Kurt Binder and Dieter~W Heermann.
\newblock {\em Monte Carlo simulation in statistical physics: an introduction}.
\newblock Springer, 2010.

\bibitem{tolman}
R.~C. Tolman.
\newblock {\em The Principles of Statistical Mechanics}.
\newblock Oxford University Press, New York, 1938.

\bibitem{farquhar}
I.~E. Farquhar.
\newblock {\em Ergodic Theory in Statistical Mechanics}.
\newblock Interscience Publishers, Inc, New York, 1964.

\bibitem{dorfman99a}
Jay~Robert Dorfman.
\newblock {\em An introduction to chaos in nonequilibrium statistical
  mechanics}.
\newblock Number~14. Cambridge University Press, 1999.

\bibitem{mountain89me}
Raymond~D Mountain and D~Thirumalai.
\newblock Measures of effective ergodic convergence in liquids.
\newblock {\em The Journal of Physical Chemistry}, 93(19):6975--6979, 1989.

\bibitem{de2005diagnosing}
Vanessa~K de~Souza and David~J Wales.
\newblock Diagnosing broken ergodicity using an energy fluctuation metric.
\newblock {\em The Journal of Chemical Physics}, 123(13):134504, 2005.

\bibitem{neirotti2000approach}
JP~Neirotti, David~L Freeman, and JD~Doll.
\newblock Approach to ergodicity in monte carlo simulations.
\newblock {\em Physical Review E}, 62(5):7445, 2000.

\bibitem{tiampo03a}
KF~Tiampo, JB~Rundle, W~Klein, JS~S{\'a} Martins, and CD~Ferguson.
\newblock Ergodic dynamics in a natural threshold system.
\newblock {\em Physical Review Letters}, 91(23):238501--238501, 2003.

\bibitem{tiampo2007a}
KF~Tiampo, JB~Rundle, W~Klein, J~Holliday, JS~S{\'a} Martins, and CD~Ferguson.
\newblock Ergodicity in natural earthquake fault networks.
\newblock {\em Physical Review E}, 75(6):066107, 2007.

\bibitem{peters11a}
Ole Peters.
\newblock Optimal leverage from non-ergodicity.
\newblock {\em Quantitative Finance}, 11(11):1593--1602, 2011.

\bibitem{thirumalai1989ergodic}
D.~Thirumalai, R.D. Mountain, and T.R. Kirkpatrick.
\newblock Ergodic behavior in supercooled liquids and in glasses.
\newblock {\em Physical Review A}, 39(7):3563, 1989.

\bibitem{hopfield1982}
John~J Hopfield.
\newblock Neural networks and physical systems with emergent collective
  computational abilities.
\newblock {\em Proceedings of the national academy of sciences},
  79(8):2554--2558, 1982.

\bibitem{compagner91}
Aaldert Compagner.
\newblock Definition of randomness.
\newblock {\em American Journal of Physics}, 59(8):700, 1991.

\bibitem{janke2002pseudo}
Wolfhard Janke.
\newblock Pseudo random numbers: Generation and quality checks.
\newblock {\em Lecture Notes John von Neumann Institute for Computing}, 10:447,
  2002.

\bibitem{matsumoto98a}
M.~Matsumoto and T.~Nishimura.
\newblock Mersenne twister: A 623-dimensionally equidistributed uniform
  pseudo-random number generator.
\newblock {\em ACM Transactions on Modeling and Computer Simulation},
  8(1):3--30, 1998.

\bibitem{mastatistical}
SK~Ma.
\newblock {\em Statistical mechanics}.
\newblock World Scientific, Singapore, 1985.

\bibitem{kingman61}
JFC Kingman.
\newblock The ergodic behaviour of random walks.
\newblock {\em Biometrika}, pages 391--396, 1961.

\bibitem{pakes69}
AG~Pakes.
\newblock Some conditions for ergodicity and recurrence of markov chains.
\newblock {\em Operations Research}, 17(6):1058--1061, 1969.

\bibitem{isingLenzMC}
M.~S\"uzen.
\newblock {\em isingLenzMC: Monte Carlo for classical Ising Model}, 2014.
\newblock v0.2
  \url{http://cran.r-project.org/web/packages/isingLenzMC/index.html}.

\end{thebibliography}
\end{document}